\ifnum\count101=1

\documentclass[12pt,twoside]{article}   
\usepackage{epsfig,times}  
\usepackage{color}

\begin{document}
\thispagestyle{empty} 

\markboth{{ \sl \hfill Chapter 1 \hfill \ }}
         {{ \sl \hfill Introduction \hfill \ }}

 \renewcommand{\topfraction}{.99}      
 \renewcommand{\bottomfraction}{.99} 
 \renewcommand{\textfraction}{.0}


\newcommand{\nc}{\newcommand}


\nc{\qI}[1]{\parindent=0mm \vskip 8mm 
{\centerline{\LARGE \color{red}#1}}\vskip 3mm}
\nc{\qA}[1]{\vskip 2.5mm \noindent {{\bf  \color{blue}  #1}} \vskip 1mm
 \parindent=0mm}
\nc{\qun}[1]{\vskip 1mm \noindent {\sl \color{blue} #1 }\quad }
%

\def\qbu{\hfill \par \hskip 6mm $ \bullet $ \hskip 2mm}
\def\qee#1{\hfill \par \hskip 6mm #1 \hskip 2 mm}

\nc{\qfoot}[1]{\footnote{{#1}}}
\def\qL{\hfill \break}
\def\qpar{\vskip 2mm plus 0.2mm minus 0.2mm}
\def\tvi{\vrule height 12pt depth 5pt width 0pt}
\def\qtvi{\vrule height 2pt depth 5pt width 0pt}
\def\qth{\vrule height 15pt depth 0pt width 0pt}
\def\qtb{\vrule height 0pt depth 5pt width 0pt}

\def\qparr{ \vskip 1.0mm plus 0.2mm minus 0.2mm \hangindent=10mm
\hangafter=1}

\def\qdec#1{\par {\leftskip=2cm {#1} \par}}

\def\qdpt{\partial_t}
\def\qdpx{\partial_x}
\def\qddpt{\partial^{2}_{t^2}}
\def\qddpx{\partial^{2}_{x^2}}
\def\qn#1{\eqno \hbox{(#1)}}
\def\qds{\displaystyle}
\def\qw{\widetilde}
\def\qmax{\mathop{\rm Max}}   
\def\qmin{\mathop{\rm Min}}   

\def\qv{\vskip 0.1mm plus 0.05mm minus 0.05mm}
\def\qhu{\hskip 1mm}
\def\qhv{\hskip 3mm}
\def\qvv{\vskip 0.5mm plus 0.2mm minus 0.2mm}
\def\qhw{\hskip 1.5mm}
\def\qleg#1#2#3{\noindent {\bf \small #1\qhw}{\small #2\qhw}{\it \small #3}\qv }


\color{yellow} 
\hrule height 50mm depth 5mm width 180mm 
\color{black}
\vskip -50mm
\centerline{\bf \Large \color{blue} How can population pyramids be used}
\vskip 4mm
\centerline{\bf \Large \color{blue} to explore the past?}    

\vskip 5mm
\centerline{\bf Bertrand M. Roehner$ ^1 $}

\vskip 4mm
\centerline{University of Paris (UPMC) and Beijing Normal University}

\vskip 3mm
\centerline{\bf 1 October 2010}
\vskip 15mm

\large

{\color{blue} \large \bf Abstract}\qL
In the same way as the rings of trees give us useful information
about the climate many decades ago (or even centuries in the
case of big trees), population pyramids allow us to know
birth or death rates several decades earlier. 
Naturally, they can fulfill such promises only if they have
been recorded accurately. That is why we start this study
by listing a number of pitfalls that may occur in 
the pricess of taking censuses.\qL
In this paper our main goal is to show how it is possible
to ``read'' population pyramids. 
Sudden changes in birth rate give rise to
the clearest signatures. This indicator reveals
that major tragedies like famines, wars, or
epidemies are generally accompagnied by a fall in birth rates.
Thanks to this observation, population pyramids
can be used to identify and gauge the blows suffered by
nations.\qL
As an illustration of how this works, we compare the
population pyramids of North and South Korea. 
Among other things, this
allows us to gauge the scale of the food shortage
that North Korea experienced at the end of the 1990s.

\vskip 1cm

\vskip 1cm

{\normalsize
1: Institute for Theoretical and High Energy Physics (LPTHE),
University Pierre and Marie Curie, Sorbonne Universit\'e,
Centre de la Recherche Scientifique (CNRS).
Paris, France. \qL
Email: roehner@lpthe.jussieu.fr}

\vfill\eject

\fi

\ifnum\count101=0
\null\vskip 1cm

\def\qvr{\color{red} \vrule height 7.3mm depth 2mm width 4pt}

$\vbox{\offinterlineskip
\halign{
\qvr \hskip 5mm#& # \cr
& {\LARGE \bf Chapter \ {\color{red} \timesgb 1} \hfill\hskip 50mm}\cr
& {\LARGE \bf How to use population pyramids to explore the past \hfill}\cr
}}$ \hfill
\thispagestyle{empty}

\fi

\vskip 2cm

\qI{Why should econophysicists analyze past events?}

Before coming to our topic we should perhaps answer an obvious question
that is related to our title. Why should econophysicists care about the
past? The answer is simple. The past is the {\it laboratory} which
allows us to test our understanding of social phenomena.
This answer may perhaps seem surprising at first sight.
It can best be explained through a parallel with astrophysics. 
\qpar

Let us consider an astrophysicist who has built a
theory of triple star systems that he wants to test. What will he do?
His first step will be to identify a sample of triple star systems
in a star catalogue, for instance the Hipparcos or the PPM
star catalogue. In a second step he will perform on these systems
the observations that are required for testing the theory. 
Likewise an econophysicist who wants to test a theory of 
(for instance) peasant
uprisings will first identify a number of such events in the past history
of countries with which he is familiar. In a second step he will need
some reliable quantitative data about these events. Most often
peasant uprisings have been militarily defeated by the armies of the
government against which the rebellion was initially directed%
\qfoot{The outcome of the civil
war in China (1927-1949) was one of a handful of exceptions.}%
.
The numbers of people killed during and after such
conflicts provide rough estimates of the magnitude of such
phenomena. 
\qpar

In other words, provided they contain some reliable data, history books
serve the same purpose for econophysics as star catalogues for
astrophysicists. This point now begins to be well understood.
It is probably for this kind of reason that the last book written
by A.-L. Barb\'asi (2010), a well-known expert in network science, 
contains a meticulous description of a 
peasant uprising that occurred in the 16th century in Transylvania (now in
Rumania) which is Barab\'asi's home region. To be sure, the description of 
just {\it one} uprising is far from constituting a catalogue of uprisings
but one may expect that Barab\'asi's very active group will produce such
a catalogue within the next years.
\qpar

In physics and in astronomy, every time a new and more accurate
observation device has been invented it has led to major progress.
Galileo's telescope constitutes one of the early example.
More recently, the invention of the multiwire proportional
chamber by the physicist Georges Charpak was a major step forward
in experimental particle physics as recognized by a Nobel award
in 1992.
\qpar

To improve the accuracy of observations about births and deaths
in major historical events is the central objective of this paper.
What is presently the accuracy of such data? A specific example will
convince us that this accuracy indeed needs to be improved.
\qpar
How many civilians died in Iraq through violent death
during the invasion of 2003 and in the 4 years afterward?
A comparison of three estimates obtained
through different surveys and methodologies reveals huge discrepancies.
The total numbers of deaths were found to be equal to 600,000, 200,000
and 50,000 respectively (Browstein et al 2008, p. 446). What is 
even worse is the fact that a comparison of the methods which were
used by these study groups does not reveal why these estimates are
so different. The authors observe that there is an ongoing discussion
and that convincing arguments have been put forward which suggest
that the highest estimate may either overestimate or underestimate
the real death toll. 
\qpar
How can an analysis based on population pyramids be of some help
in such a situation? Because it relies on data for different 
age-groups such an analysis allows us to separate different sorts
of death such as infant mortality, death of adult males, old age
mortality. Once the mortality in a ``normal'' population has
been discounted one will be able to identify and estimate
the excess-mortality due to the occupation (whether by
violent deaths or not)
\qfoot{Naturally, population pyramid data must be accompanied
by data about emigration because population pyramids
make no distinction between {\it permanent} emigration and death.}%
.
\qpar

The paper is organized as follows. First we recall the
definition of a population pyramid and introduce the important
distinction between static and dynamic analysis.
Before beginning to use population pyramid data we discuss
their reliability and accuracy. From a physicist's perspective
this is a crucial step.
Then, we explain how population pyramids allow us to measure birth 
rates. The response of birth rates to several kinds of ``special events'' 
is illustrated through various examples.
Finally, we propose an agenda for further research.
An appendix which provides useful
information about population pyramid and census databases
closes the paper.

\qI{How can population pyramids serve to explore the past?}

A population pyramid gives the structure by age and sex of a population.
Various examples are given in Fig. 3a,b,c,d and Fig. 5.

\qA{A spatial parallel of aging}
In physics one is more used to movement in space and time than to
aging processes. So it may be worthwhile
to describe a spatial parallel of the aging process
which may be more suggestive to physicists.
Fig. 1 pictures a parallel in which age
has been replaced by a spatial variable. 
\qpar
Let us imagine that 
from a bridge on which she is standing a young girl
drops small paper boats into a river.
%
\begin{figure}[tb]
    \centerline{\psfig{width=10cm,figure=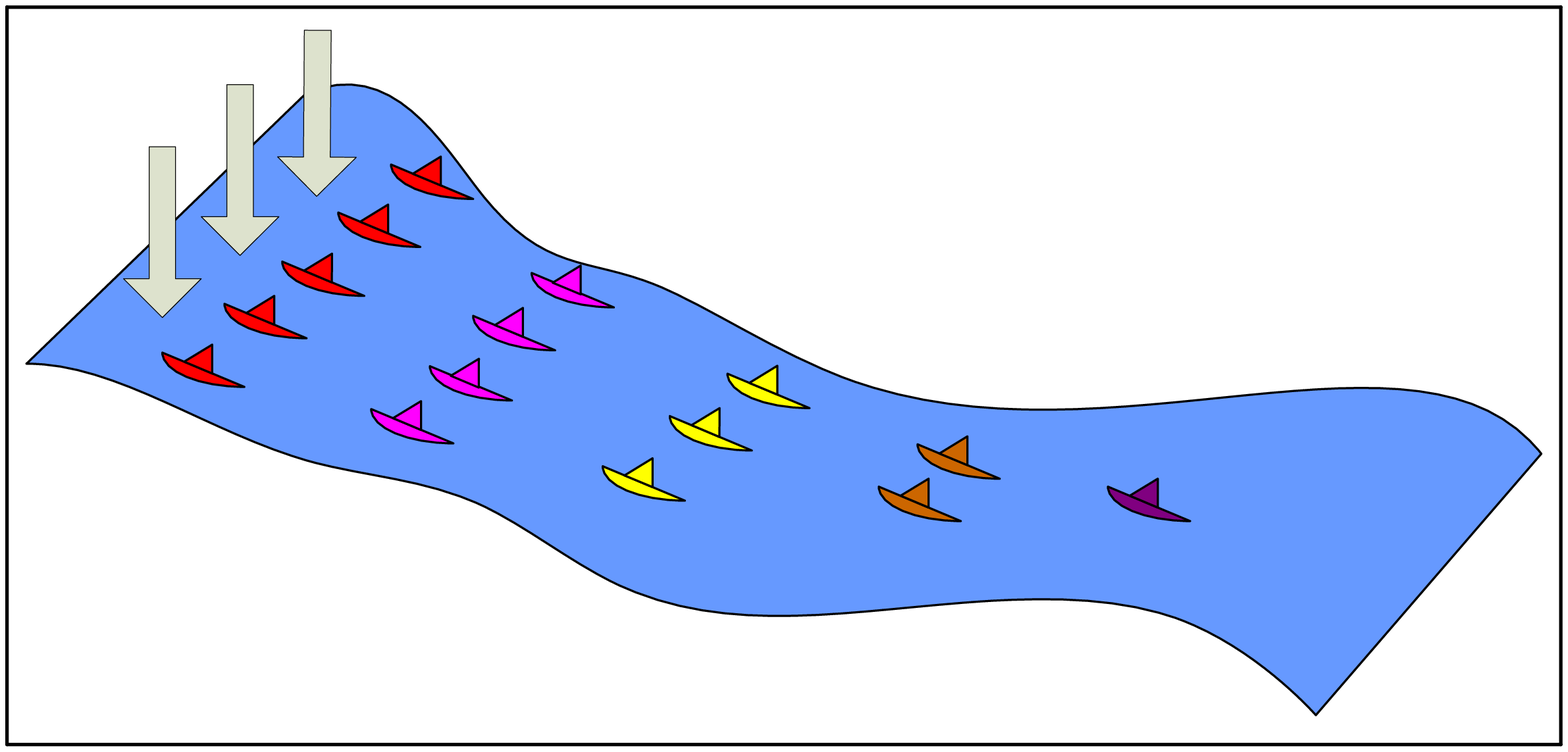}}
\vskip 4mm
\qleg{Fig.\qhu 1:\qhv Spatial analogy of the process of aging.}
{The blue color represents the water of a river. At one point
shown by the vertical arrows, small paper boats are dropped
into the water in several successive batches.
Each batch (represented by a different color) has
approximately the same number of boats except for some random
fluctuations which represent the more or less random fluctuations
of birth rates.
Once in the water, the boats begin to move
downstream but as a result of various accidents their number
decreases in the course of time until eventually all of them have
sunk.
In demography, each row of boats 
is called a cohort. The observation of one specific row in its movement
downstream is called cohort analysis. Such an analysis should 
be distinguished from the static, instantaneous
picture provided by population pyramids.}
{}
\end{figure}
%
At 10am she drops a batch of
25 boats of red color. At 10:10 she drops a second batch consisting of
28 blue boats, at 10:20 she drops a third batch of 21 boats
and so on with
each batch of boats corresponding to a different color. 
Once in the water, the paper boats are carried downstream by the
current. At intervals some of the boat sink, either because
affected by a whirlpool or perhaps because they landed in the water 
on the side and quickly took in water. This analogy will
help us to explain a distinction which plays an important role
in demography.
\qpar

The aging process of a population can be analyzed in two different
ways, static or dynamic. In demography the dynamic viewpoint
is called ``cohort analysis''.
This important
distinction has a clear interpretation in the paper boat analogy.
\qbu The static viewpoint corresponds to population pyramids.
At a given moment one takes a picture of the flotilla of paperboats
and one counts the number of boats of each color. It should be noted that
between two successive rows there is
no clear connection because the numbers in successive batches are
fairly random.
\qbu In the cohort analysis one concentrates on a specific
batch as it moves down the stream. 
Whereas the static view is an instantaneous 
snapshot, the cohort viewpoint requires an
observation over a long time interval from the moment when the 
boats fall together in the water to the moment when the last boat
of this batch
sinks; in demography such an observation must cover over
one century. Instead of the expression ``cohort analysis''
which is fairly abstract we will rather use the expressions
``downstream''  or``along the stream'' analysis
which make reference to the river analogy.

\qA{Stability of downstream changes versus instability of birth rates}

As we will see later on, birth rates are highly fluctuating.
Wars, periods of high food prices, 
displacements of population may bring about a reduction in birth rates.
On the contrary, after the end of wars or during periods of
economic prosperity birth rates will tend to increase. 
In parallel with births, babies and small children
who are much more vulnerable than adults to
external events will also be affected to a greater degree.
\qpar
Apart from this and
for reasons we do not yet understand,
there are dramatic changes in fertility rates (i.e.
number of children per woman) which have strong effects on
birth rates. For instance, between 1990 and 2002, Poland 
experienced a dramatic reduction in its fertility rate
from 2.0 to 1.35. Similar changes occurred 
during the same decade in several other
countries such as for instance Italy, South Korea, Spain or
Ukraine.
\qpar
It is for these different reasons that in our
stream analogy the number of boats in successive batches was
taken as a random variable. 
\qpar

On the contrary, once a generation has reached the age of 5
its reduction rate in the course of time
will be fairly stable, only determined
by the overall standard of living and availability of 
healthcare. In short, when we follow a given age group
downstream we would expect a very smooth and predictable evolution.
However, there are two kinds of circumstances which will 
disrupt this smooth evolution.
\qbu Unexpected events which produce mass mortality such as
wars, epidemics or earthquakes. It is precisely in such events
that we are interested. It is because they will appear as
sharp falls on a fairly smooth curve that they can be identified
and measured.
\qbu When we follow an age-group downstream mass emigration
or immigration will also produce rapid changes. 
In fact, the data 
make no distinction between people who died or people who
moved away. In both cases, they just disappear
from the census statistics.

\qI{What is the reliability and accuracy of census data?}

It is a distinctive feature of physicists to care about the
quality of the data that they use. Social scientists and
particularly economists do not. This can be illustrated by two
facts: (i) Even though the accuracy of data is
much higher in physics than in the social sciences, physicists
publish them with error bars whereas social scientists do not.
(ii) By browsing through papers published in economic journals
one quickly comes to realize that usually
only 2 or 3 lines are devoted to
discussing the origin, reliability and accuracy of the data
that will be used. 
Yet, if the data are flawed the analysis will be faulty, 
confusing and useless. This makes
the question of data accuracy a matter of cardinal importance.
That is why we will give close attention
to the various pitfalls that may occur.
\qpar

Population pyramids rely on census data. 
Among statistical data
recorded by government agencies, census data are considered
as particularly reliable and accurate. There are two main
reasons for this.
\qbu Because censuses were among the first operations 
that were carried out, statistical agencies were able to develop
adequate procedures and improve them in the course of time.
\qbu In contrast to macroeconomic data (e.g. statistics
of Gross Domestic Product)
which require heterogeneous variables to be aggregated
(a process often referred to as the problem of
``adding apple to oranges'') 
population statistics only involve the addition of
data of a single kind.
\qpar
In spite of this, 
population data (as indeed any data based on observation)
should be considered carefully and viewed with a critical eye.
\qpar
Basically, population
pyramids by country and province 
require information about residence and age.
Are these variables easy to define and to collect?

\qA{Residence}
As is well known, the specification of the residence 
poses some problems for certain categories of citizens who 
move frequently such as for instance the personnel of the navy
or army. Such problems can be solved by setting up
clear rules and definitions. \qL
Let me give an illustration. At first sight,
the big dip
in the male population shown in  Fig. 2 may
look surprising. It becomes clear
once one knows that in the 1990s the North Korean
government has made the choice of counting the military personnel
separately from the rest of the population%
\qfoot{Most countries do not include in their censuses the personnel
of armed forces who are located overseas; counting apart
the military based in the country is more uncommon.}%
.

\begin{figure}[tb]
   \centerline{\psfig{width=10cm,figure=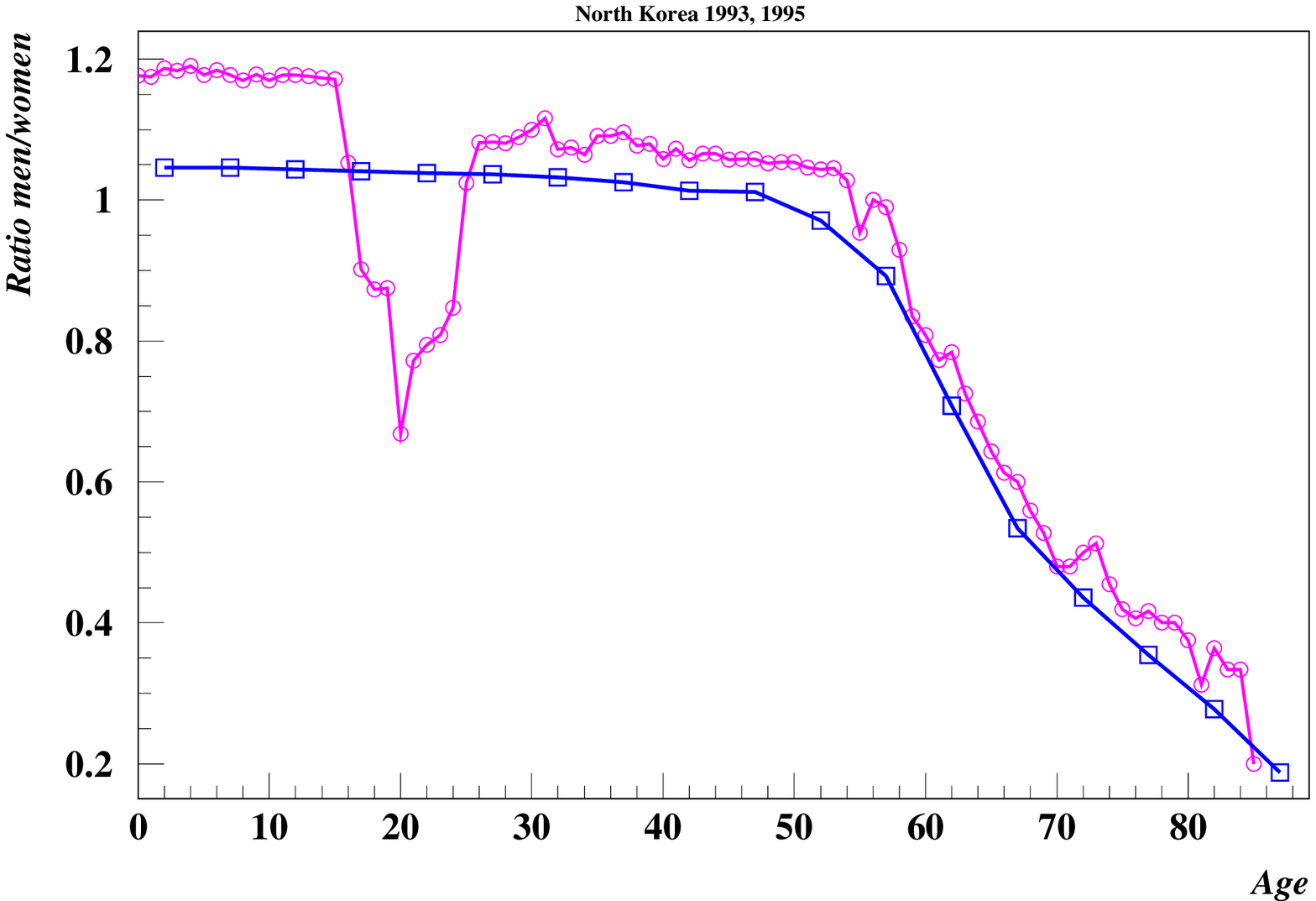}}
\qleg{Fig.\qhu 2:\qhv Sex ratio by age in North Korea.}
{The big dip in one of the curve is due to the fact that 
the military have not been included in the census of 1993.
The blue curve which refers to 5-year age-groups has been 
corrected for this effect on the basis of previous censuses.
At this point it is not clear why it gives a ratio
which is lower than the other curve in ages from 0 to 18.}
{Sources: United Nations: population by single years of age;
United Nations: World Population Prospects, 2008 revision.}
\end{figure}

\qA{Age}
One may think that for age there should be less problems
than for place of residence.
Observation shows otherwise. This is can be illustrated by 
the following examples.
\qbu The population pyramid of France based on the census of 1851
(Fig. 3a)
shows big steps for the ages of 30, 40, 50, 60 and 70 years.
For instance the population for the 50-year age group is almost
twice the populations of the 49- or 51-year age groups.
Why is this so? \qL
A comparison with population pyramids of the same period
in other countries such as Canada or the United States
(especially at state level%
\qfoot{For instance at the census of 1850 in Georgia
the age-group of 30 has a size four times larger than those
for 29 or 31.}%
) 
shows the same feature. On the
contrary, twentieth century data do {\it not} exhibit such a feature.
The most likely explanation is that this feature comes from
the recording procedure. First, one must realize that French
people who were 50 in 1851 were born in 1801 at a time when 
the registration of births by state agencies was just beginning.
As a result, many of them did not know their age
with precision. To the question ``How old are you'' asked
by the census officer, they may have answered ``Around 50''
which was then recorded as 50 on the census form. In short,
by giving the closest round number as a proxy of their age
people favored multiples of 10 but also 
(albeit to a smaller extent) multiples of 5.
\qbu In the case of India (Fig. 3b), in addition to the multiples of 10,
the steps for ages which end with 2 or 8 are also larger than
expected. The number 8 is considered as a ``good'' 
number in China. Is it the same in India? We do not know.
\qbu As mentioned above,
the bias in favor of multiples of 10 and 5 also occurs
for the 1850 census in the US state
of Georgia (Fig. 3c). In this case there are in addition substantial
random fluctuations due to the fact that the pyramid is based on a 1\% 
sample of the total population. 
\qbu The case of Chile (Fig. 3d)
is somewhat different in the sense that in
this case the steps which are enlarged are those which end with 2.
The age groups for 32, 42, 52, 62, 72 have a size which 
is systematically larger than the adjacent age groups. The difference
is not as big as in the previous examples but it is nevertheless
significant. For instance the 42-year age group is 30\% larger than
the 41 or 43 age-groups. How can one explain this feature? 
This can possibly be explained by assuming that the census agent
did not ask the age but the year of birth. 
Thus, because the census was done in 1992, 
if a person
born {\it around} 1950 rounds the number to 1950 that
would produce an excess of people who were 42 year old in 1992.


\begin{figure}[t]
\setbox41=\vbox{ \hsize=8cm
\centerline{\psfig{width=7.5cm,figure=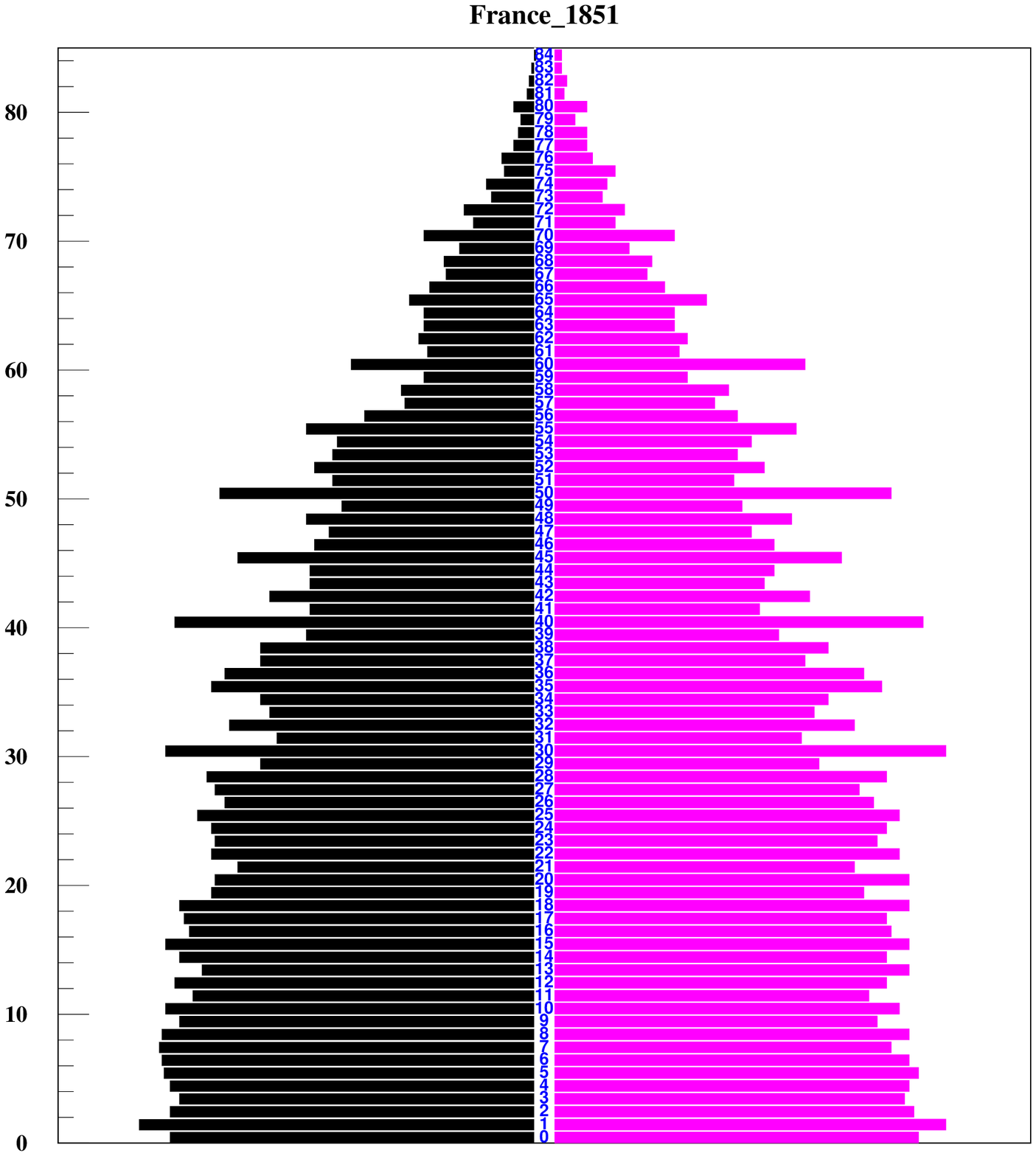}} 
\vskip 2mm
{\bf \color{blue} Fig. 3a: Population pyramid of France in 1851.}
{\color{blue} The steps of ages that are multiples of 10 are
larger than they should be.}
{\it \color{blue} Source: Annuaire statistique de la France, 
Annuaire r\'etrospectif (1966)}
}
\setbox42=\vbox{ \hsize=8cm
\centerline{\psfig{width=7.5cm,figure=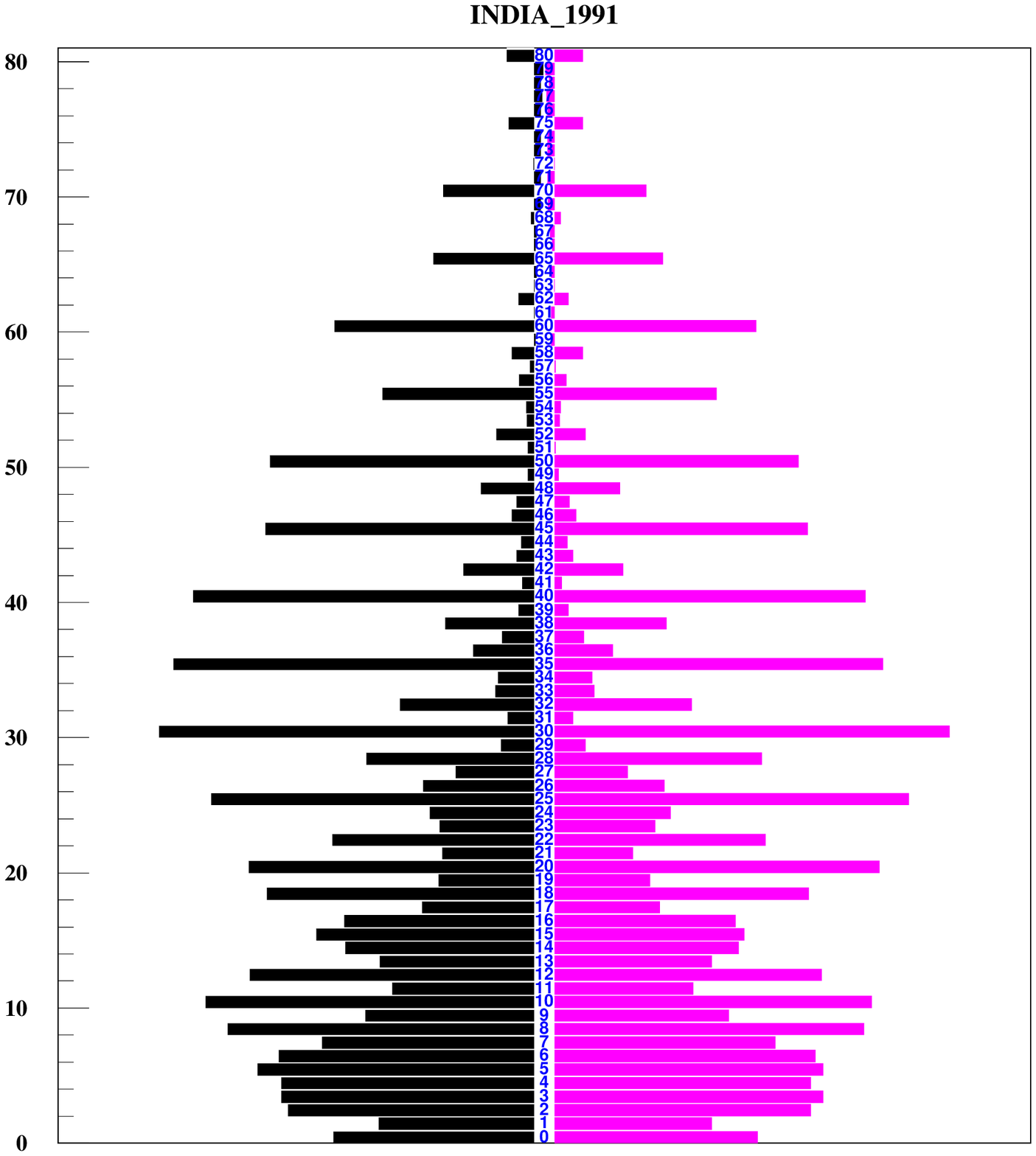}} 
\vskip 2mm
{\bf \color{blue} Fig. 3b: Population pyramid of India in 1991.}
{\color{blue} The steps ages which are multiples of 10 or which end
with 2 and 8 are much larger than expected.}
{\it \color{blue} Source: United Nations: Population by single 
years of age.}
}
\centerline{\hfill \box41 \hskip 2mm \hfill \hskip 2mm \hfill \box42 \hfill}
\end{figure}
%
\vskip 3mm
%
\begin{figure}[h]
\setbox41=\vbox{ \hsize=8cm
\centerline{\psfig{width=7.5cm,figure=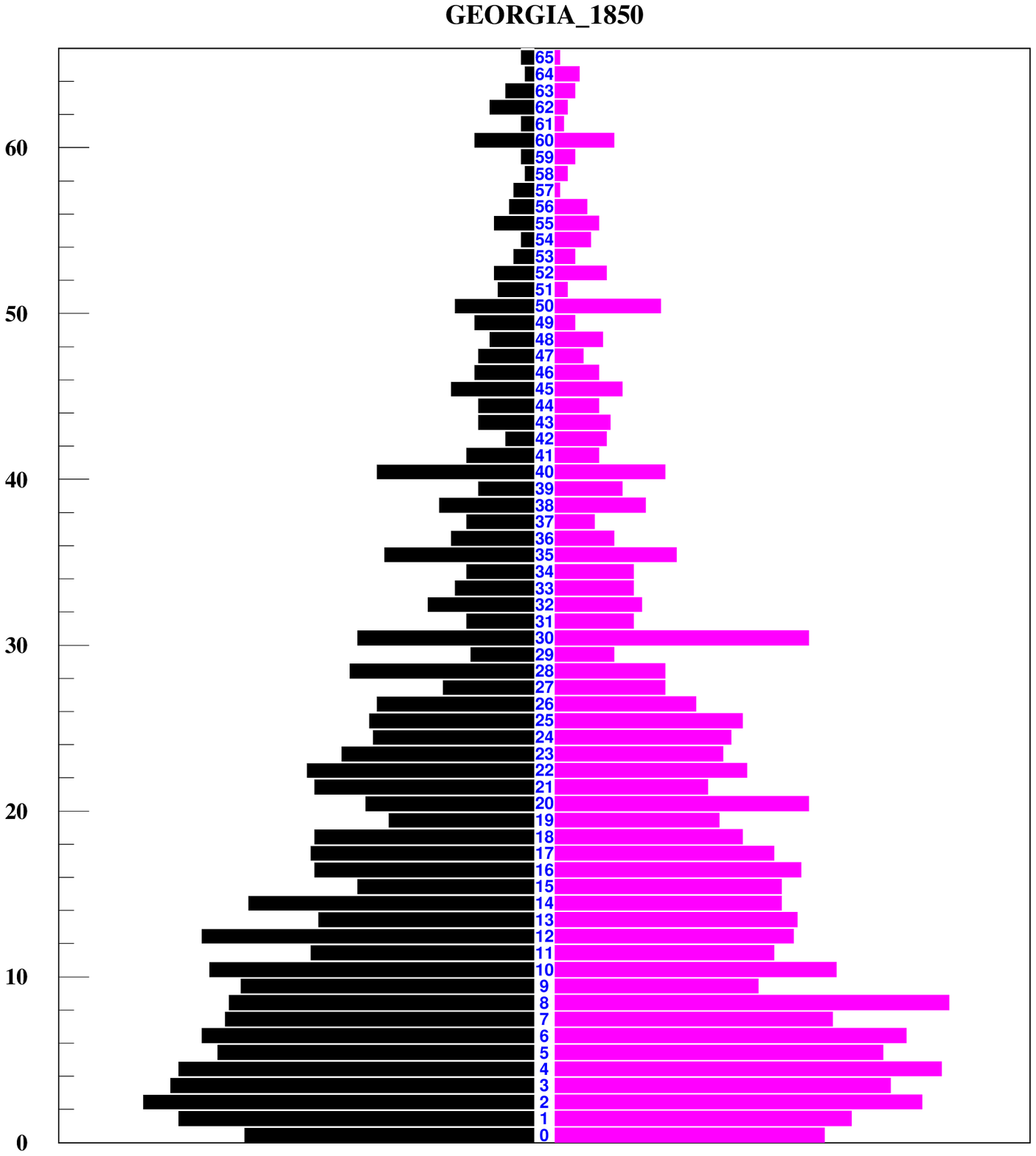}} 
\vskip 2mm
{\bf \color{blue} Fig. 3c Population pyramid of the US state of
Georgia in 1850.} 
{\color{blue} The steps of ages that are multiples of 10 are
larger than expected.}
{\it \color{blue} Source: IPUMS (USA).} 
}
\setbox42=\vbox{ \hsize=8cm
\centerline{\psfig{width=7.5cm,figure=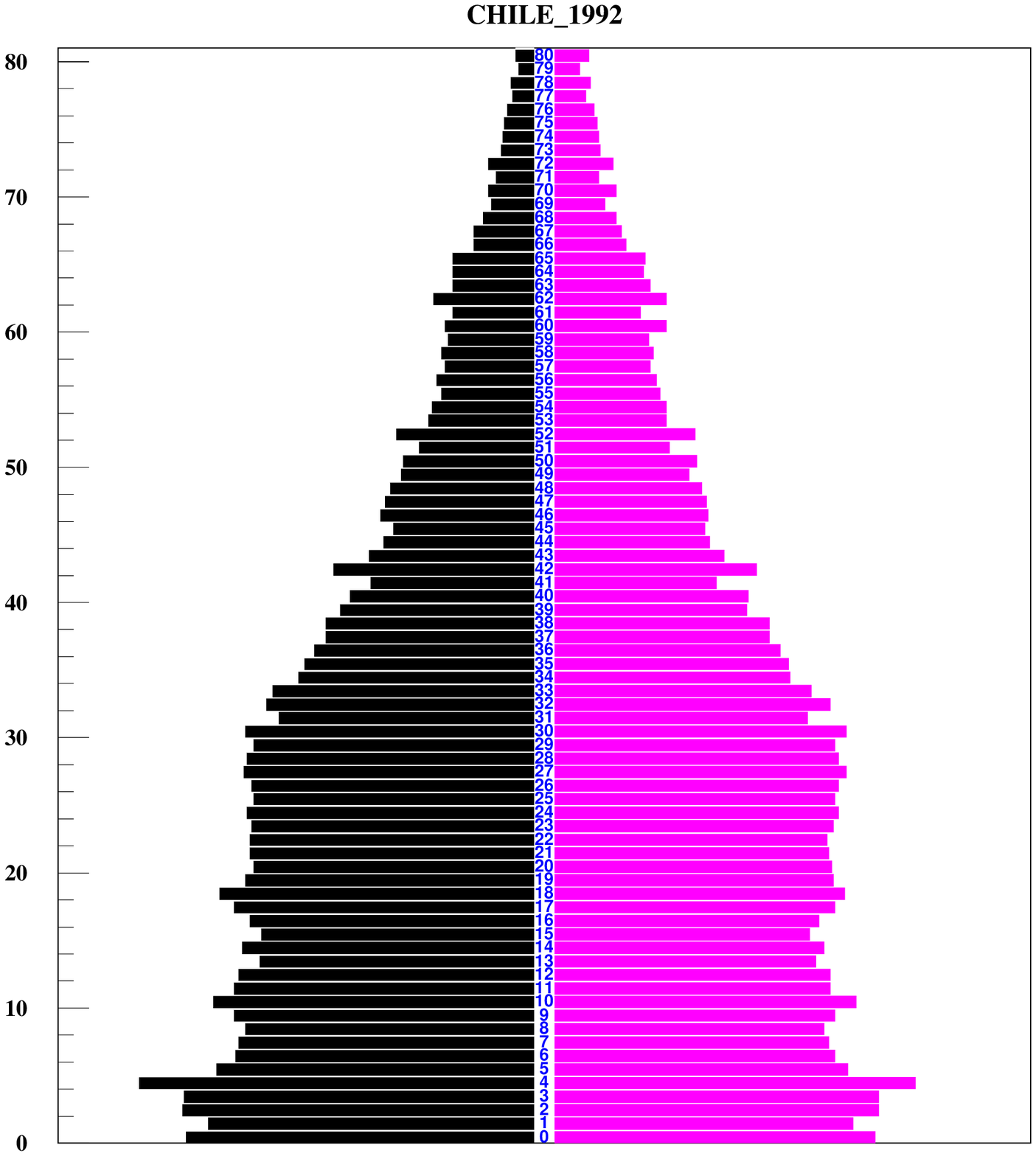}} 
\vskip 2mm
{\bf \color{blue} Fig. 3d: Population pyramid of Chile in 1992.}
{\color{blue} the steps of ages 32, 42, 52, 62, 72 are larger
than expected.}
{\it \color{blue} Source: United Nations: Population 
by single years of age.} 
}
\centerline{\hfill \box41 \hskip 2mm \hfill \hskip 2mm \box42 \hfill}
\end{figure}

\qpar

Whatever the exact reasons of these discrepancies, they are
very damageable for any precise analysis. To some extent they can be 
eliminated by constructing broader age-groups for the 
following age-intervals:
$$ (0.5,10.5)\ (10.5,20.5)\ (20.5,30.5) \ldots $$

In this way, 
one has age-groups of equal amplitude and the
big steps at $ 10,\ 20,\ 30, \ldots $ will be divided uniformly
between the two adjacent decades. Of course, such a treatment
will probably not be sufficient to make a very irregular pyramid
(such as the one for India) acceptable.

\qI{How to estimate birth numbers from population pyramids?}

Birth and death rates are the two main demographic variables.
Both can be estimated by using the information contained in
population pyramids but birth rates are much easier to estimate
than death rates. In this section we discuss how this can be done
and how reliable such estimates are.
\qpar

The first step of a 
population pyramid for year 1995 (for instance) gives
of course the numbers of birth in this year. Does 
the population pyramid also
allow us to estimate birth numbers for the years before 1995
and, if so, for how many years? For instance, the people who are
20 year-old in 1995 were born in 1975; can their number be used
as a proxy for the number of people born in 1975?
The answer to this question depends on what happened between
1975 and 1995. If in those years there was a big disaster or
a disease which greatly affected the young people, then the
age-group born in 1975 will be so much reduced until it reaches
the age of 20 that it cannot well reflect its initial birth
size. In other words, there can be no general answer to our
question. It all depends on whether there have been special
circumstances or not.
\qpar

It will be helpful to illustrate this argument by an example.
Fig. 4 provides a comparison between birth numbers 
(broken line) in Japan
and age-group sizes (solid lines) derived from 3 population pyramids.

\begin{figure}[tb]
    \centerline{\psfig{width=17cm,figure=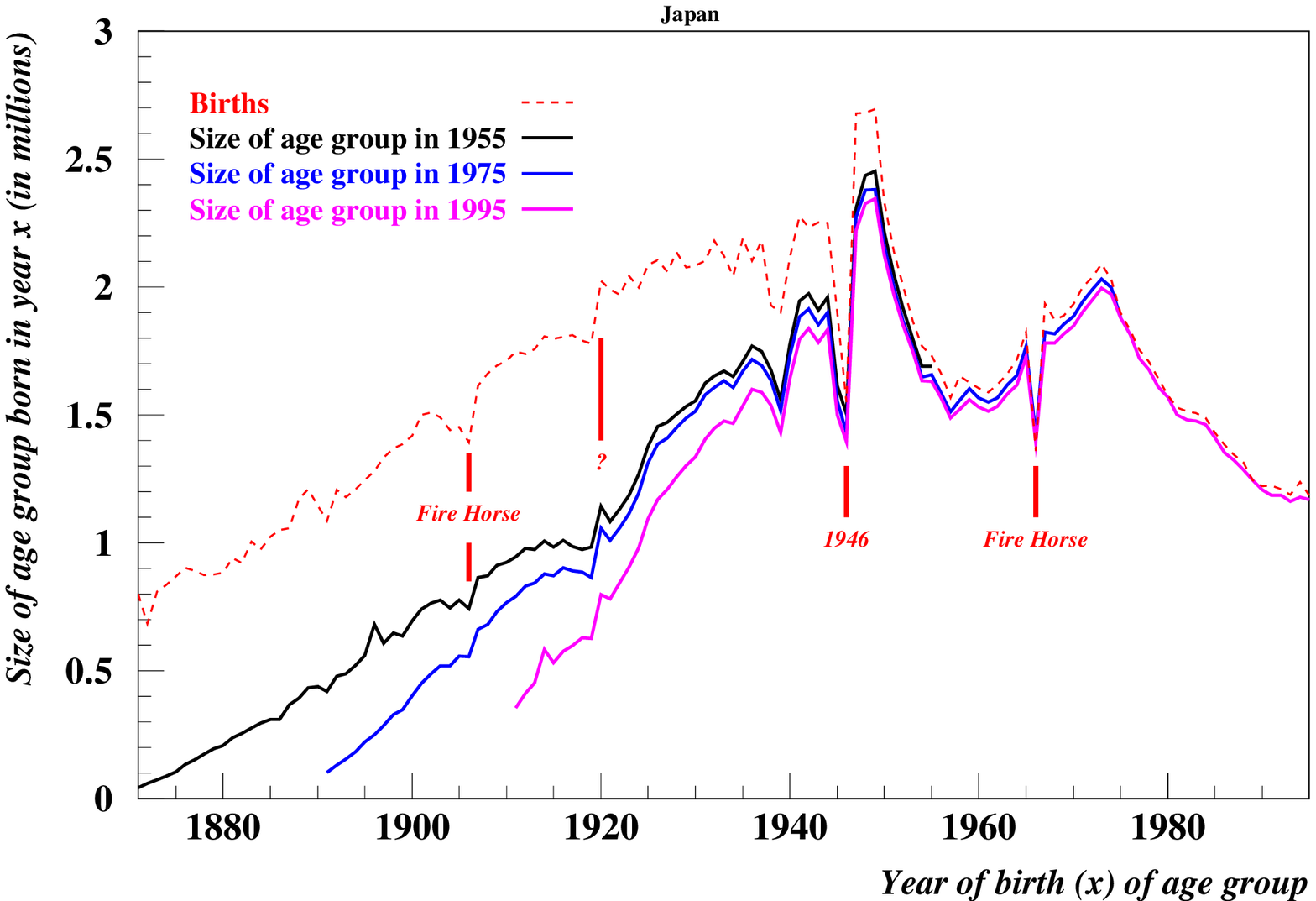}}
\qleg{Fig.\qhu 4:\qhv Comparison of birth numbers with the sizes of
steps of population pyramids.}
{The fact that the 3 population pyramid curves closely
follow the ups and downs of the number of births shows that,
unless there are special circumstances, population
pyramids provide good estimates of birth numbers.
Between 1947 and 1957 the infant mortality rate was divided
by 3, this explains the sudden increase in the distance
between the broken and solid curves in the years before 1950.
The fact that it became momentarily
smaller in 1946 can be explained in only
two ways (i) The birth data for 1946 may be incorrect (there is 
considerable uncertainty for Japanese vital rates 
in the years immediately after 1945)
(ii) There was a massive inflow of Japanese (from
Korea and Japan) with babies of less than one year.}
{Source: Historical Statistics of Japan: online database
of the Ministry of Internal Affairs and Communications. The
birth data are from Liesner (1989)}
\end{figure}

First, we must say how the 3 solid line curves were derived
from the age-group numbers given by the pyramids.
They are in fact identical to the steps of the pyramids but
instead of the age it is the birth year which is used to 
label the age-groups, For instance, the people of age 20 
in the pyramid of 1955 were
born in 1935 and in 1955 they numbered about 1.7 million. 
Similarly, the people of age 60
in the pyramid of 1995 were also born in 1935, but in 1995
their number has been reduced to 1.5 million.
The second curve is lower
than the first because some people in this age-group have died between 
1955 and 1995. 
In a general way, the solid line curves are all lower
than the broken line curve of birth numbers; the only 
exception would occur in case of a major immigration inflow.
The only year when this may have occurred in this graph is 1946
when many Japanese people established in China or Korea came back
to Japan following the end of the war.
\qpar

The fact that, despite a high mortality of young men during the
war, the broken and solid line curves are fairly close shows that
the sizes of age-groups provide good estimates of birth numbers.
This is an important rule.
\qpar

As one goes more toward the past the solid lines curves become of course
lower but it can be seen that their short-term fluctuations
still fairly well reflect the fluctuations of birth numbers.
For instance, the sudden fall in births which occurred
in 1906 is fairly well reproduced in the pyramids of 1955 and 1975.
This event as well as the 3 others marked 
by vertical lines will be discussed in the next section.

\qI{Statistical signature of hardship through abrupt birth falls}

We have observed previously that birth rates are subject to 
many fluctuations. However,
abrupt falls over a period of
one or two years followed by a return to the
level which preceded the fall most often indicate that the population
suffered some form of hardship 
as a result of events such as war, food shortage, or other disruptions.
What makes this observation of particular interest is the
fact that, as we have seen, birth rates can be 
easily measured on population pyramids.

\qA{Wars}
\qun{France} When one looks at the population 
pyramid of France in 1936
the first thing which attracts attention is a huge indentation around
the age of 20. 

\begin{figure}[tb]
    \centerline{\psfig{width=10cm,figure=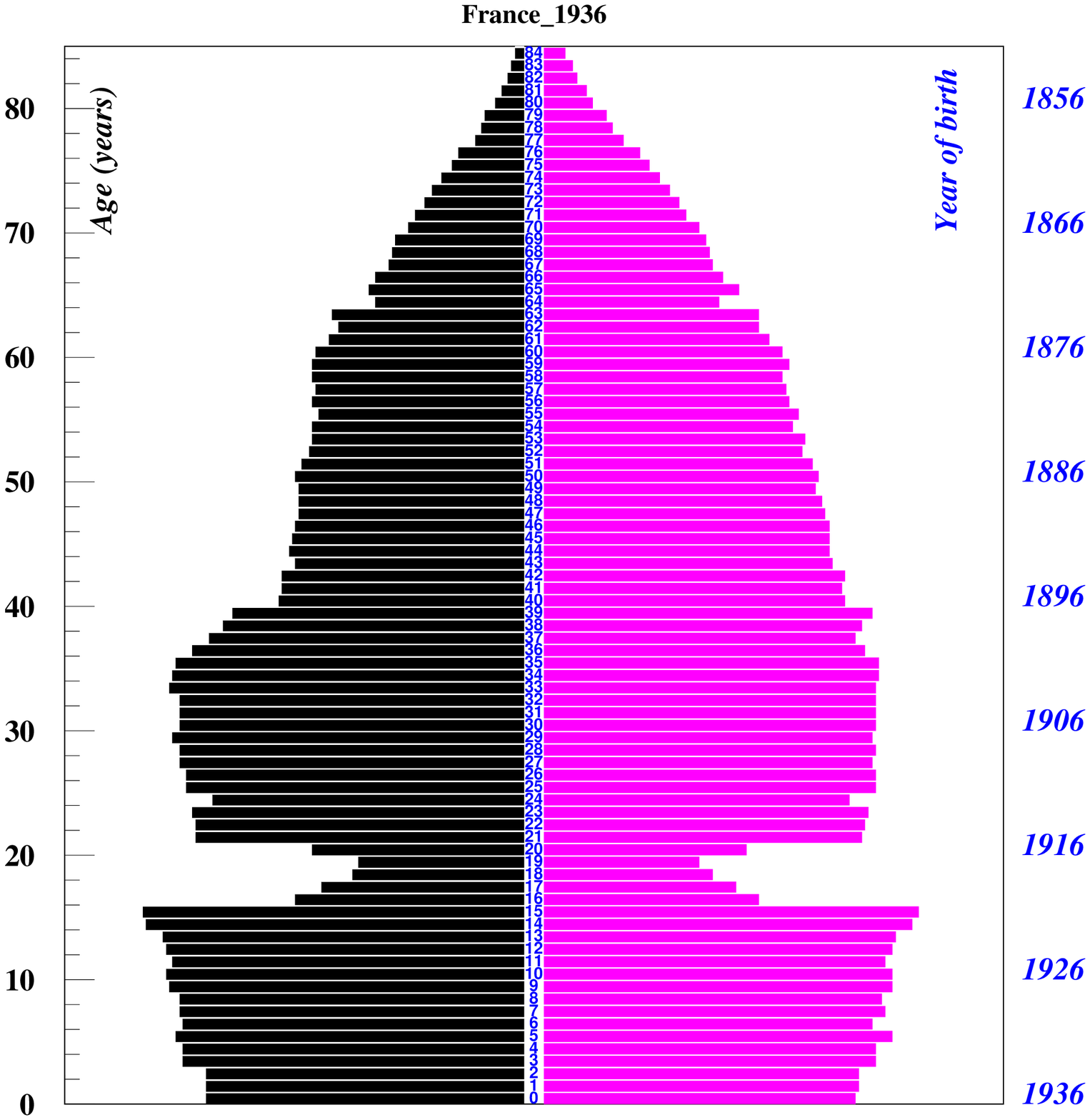}}
\qleg{Fig.\qhu 5:\qhv Reductions in births in France
during the First World War.}
{During the 5 years of
World War I France experienced a reduction in births of up to 50\%
Altogether this resulted in about one million ``non-born''
babies.
The two years after the end of the war were marked by a 
weak and short-lived babyboom; in 1920 the birth rate
was almost back to its pre-war level. In contrast, the end of World
War II was marked by a strong and lasting babyboom.
So far, we do not really understand the reason of this difference.}
{Source: Annuaire Statistique de la France 1966, R\'esum\'e R\'etrospectif.}
\end{figure}

Of course, one is tempted to make
a connection with World War I. However, this indentation does
{\it not} correspond to the soldiers who died in the war. 
As a matter of fact, the indentation is the same for males and females.
Another indication is provided by the fact that in 1936 the
people corresponding to this notch are 20-year-old which means they were
born in 1916. In short, the war brought about a massive reduction
in birth rates: they were divided by 2. 
\qpar

A natural question is whether the pyramid also allows us to identify the 
disappearance of the more than 
one million soldiers who died on the battle fields. If we look at the
step corresponding to an age of 40 (that is to say 20 in
1916) we see a reduction of 14\% in the size of the male step.
There is no such reduction on the Female side.
not present on the female side. However, this reduction 
is much less visible than the fall in birth numbers.
Not only is it much smaller but it can
be identified with certainty only because it did not affect the
women. 
\qpar
Incidentally, one can also see on this pyramid the effect of
the war of 1870-1871 between France and Prussia.
It appears as a reduction of 17\% for the step corresponding to
age 64 ($1936-64=1872 $). In this case
the deaths which occurred during the war are completely invisible
because they affected people who are 84 year old in 1936.

\qun{United States} The Civil war brought about a substantial reduction
in birth rates especially on the Confederation side. This makes sense
because in contrast to Connecticut and Massachusetts,
the four Confederate states considered in Fig. 6 were directly
confronted to military occupation by Union troops.
\qpar

In contrast, neither World War I nor World War seems to have 
led to a reduction in birth rates. How can one understand this?
For World War I, the explanation is probably that US troops
played a role only in the last months of the war.
During World War II some 12 million Americans (10\% of the population)
served in the armed forces. Thus, the lack of any reduction in
birth rate may seem more surprising. One possible explanation is the 
following. During the Great Depression there was a marked reduction
in birth rates but in 1940 with the beginning of the war came a great wave
of prosperity. The two effects probably cancelled one another.

\begin{figure}[tb]
    \centerline{\psfig{width=16cm,figure=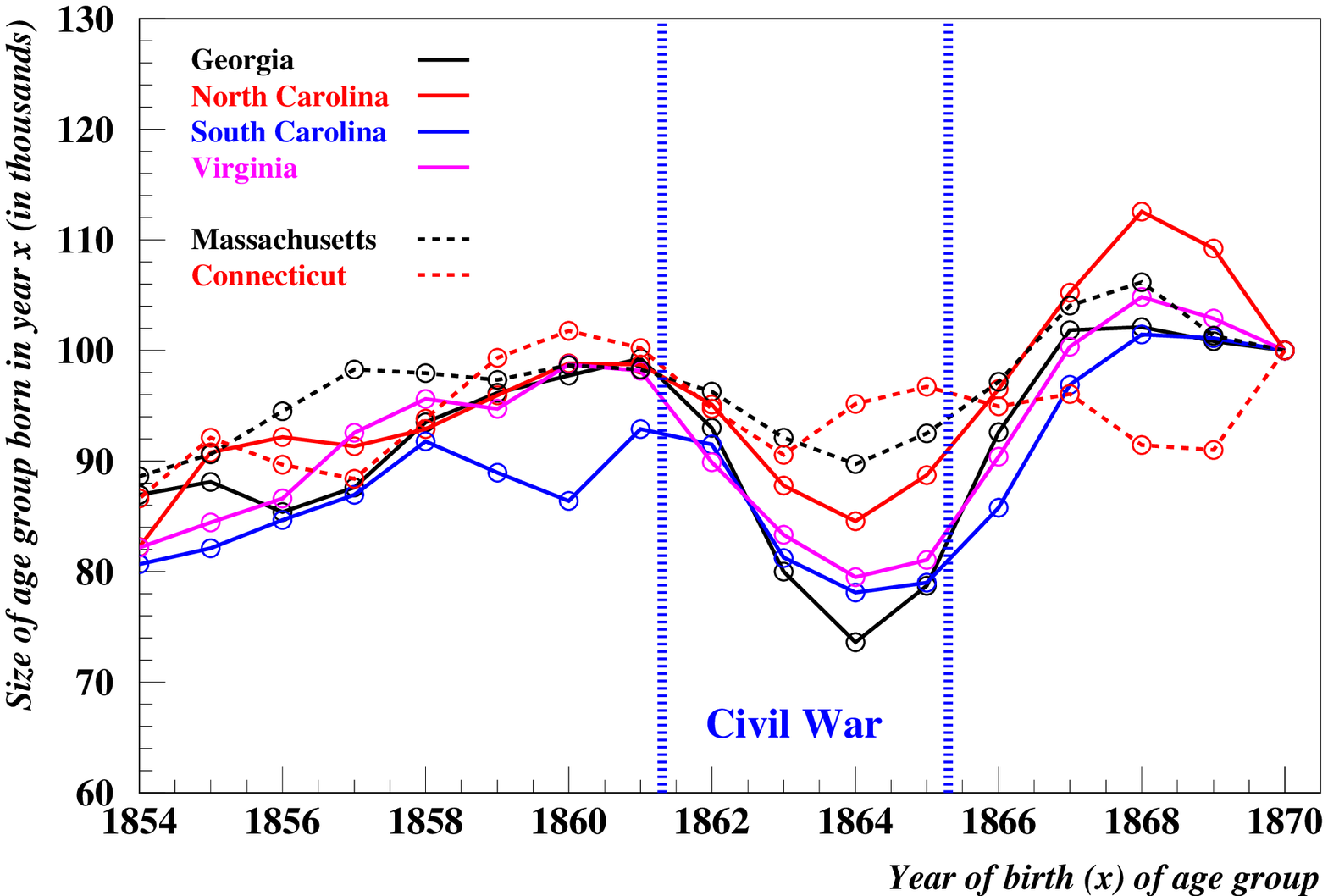}}
\qleg{Fig.\qhu 6:\qhv Reduction in birth rates during the Civil War
in the United States.}
{Solid lines correspond to Confederation states whereas broken
lines correspond to Union states.
The war was marked by falls in births of up to 30\% 
in the Confederation and of the order of 10\% 
in the Union. This suggests that, as expected, Confederate
states suffered much more than Union states. It can be noted
that no records of birth statistics are available in this period.
These states became so-called birth registration states only in the
1910s.
In other words, the only way to explore this 
demographic aspect of the Civil War is to use population pyramids.}
{Source: IPUMS USA (University of Minnesota).}
\end{figure}

\qA{Are birth rates sensitive to food shortages?}

Reductions in birth rates occur not only during wars but
also as a result of other kinds of hardships.
For instance one may wonder what is the impact of food shortages.
Nowadays food shortages are rare in developed countries
but they were fairly common in former centuries. In other
words, in order to observe the effects of food shortages 
it is a good idea to use data from former centuries. 

\qun{France between 1800 and 1850}
This was done 
by the present author in a study (Roehner 1990)
about France in the first half of the 19th century. It was
shown that in this time there was a significant correlation coefficient
($ r=-0.6 $) between the price of wheat and the number of
births. Naturally a high price of food meant scarcity and shortage for the 
poorest people. However, these reductions in birth rates were too
small (compared to the noise) to be visible on
the population pyramid of 1851.
The same study also shows that for these relatively mild food scarcities
there was {\it no} significant correlation
between wheat prices and mortality. 
\qpar

In other words, birth numbers are a much
more sensitive indicator of food shortages than death numbers.
Thereafter we will refer to this result as the 
``birth-hardship criterion''. 
\qpar

However, one must recognize that in some cases the effect on births
is surprisingly small. This is what we discuss now.

\qun{The case of Henan in 1938 and 1944}
It is well known that in the 1930s there were a number of serious
food shortages in several provinces of China. One of the most severe
occurred in Henan in June 1938 after the levees of the
Yellow River were blown up
by Kuomintang troops in a desperate attempt to stop
the Japanese Army. We are told that almost one million people
died through the flood and ensuing food shortage and that 12 million
were made homeless (see http://mygeologypage.ucdavis.edu). In
1943 there was another flood in Henan during which
(according to the same website) ``3 million of people starved to death''.
If one estimates the population of Henan in the 1940s 
to be of the order of 30 millions%
\qfoot{This rough estimate results from the population of Henan
in 2010 divided by the growth factor of the total Chinese population
which is about 3).}%
,
3 million deaths represent 10\% 
of the population. Yet, the birth rate as reflected in the
population pyramid of Henan, shows only a modest fall of 22\%
between 1941 and 1944. Between 1937 and 1939 the reduction was
even smaller, only 18\%. 
One must keep in mind the possibility 
that the figures of one and three millions
of deaths are perhaps over-estimates.
\qpar
Below we take the problem by the other end in the sense that
we consider episodes that {\it are} marked by large birth reductions. 
The problem then is to
understand what kinds of hardship were responsible.

\qun{The birth reduction of 1961 in Chinese provinces}
A clear case of birth reduction is provided by the
Chinese province of Sichuan in south-west
China. The population pyramid of this province%
\qfoot{All following results for Chinese provinces are derived from
the census of 1982, of which a 1\% sample is available
on the IPUMS International website.}
shows that
between 1958 and 1961 the number of births fell by 63\%.
In other provinces birth numbers fell also but less.
For instance, in the cities of Beijing and Shanghai the reduction
was only 30\%. 
The reduction was even smaller in the Northern provinces of
Heilongjiang, Jilin and Inner Mongolia (20\%).
The province where the reduction was the smallest was Tibet (5\%)
\qfoot{The population pyramid of Tibet is one of the smoothest and most
regular of all provinces. It has no visible indentation and moreover
the male/female ratio does not show any systematic deviation
from 1, a feature
which is in marked contrast with many other provinces.}%
.
For all provinces together the average reduction was about 40\%.
This is of the same magnitude than the reduction in France during
World War I.
\qpar


What were the factors behind these reductions? \qL
Ren\'e Dumont,
a French expert in agricultural economics
who traveled extensively throughout China in these years,
cites mainly
three factors. (i) Food shortage due to bad weather combined with
the fact that many peasants were employed far away from 
their village in the amelioration
of irrigation systems and could not take part in the harvests%
\qfoot{Dumont writes the following: ``Between 1955 and 1964
I observed the most extraordinary transformation of the agricultural
landscape. When one flies over China from Hanoi to
Beijing one sees that the regions to the south of the Yangtze
are now covered with canals, levees and dikes where only water reservoirs 
had existed previously.'' 
(Dumont 1964, p. 393, my translation)}%
.
(ii) As the men were often sent to other places 
for work on dams, 
canals or dykes, family life was disrupted. 
Moreover these rapid
changes generated a reaction of social resistance and the 
resulting conflicts added to the hardship of families
(Dumont 1964 p. 387).
(iii) Finally, there was already an attempt at 
some limited forms of birth control.
Early marriage was discouraged and late marriage encouraged;
economic advantages for children were restricted after the 
third child (Dumont 1964, p. 396).
\qpar
A confirmation of the role played by the disruption of family life
comes
from the fact that birth reductions were twice as high in regions
such as Henan and Sichuan (where hydraulic systems are
of key importance) as in the northern provinces.
\qpar

To the three factors listed by Dumont one can add a fourth that
he does not mention but which appears fairly clearly by
examining the population pyramids. After the end of the civil
war in 1950 there was a babyboom in many Chinese provinces.
Such episodes usually last only a few years. It seems
that this boom came to an end toward 1957-1958 which means 
that, together with the other factors, this contributed to 
the birth reduction. Incidentally,
after 1963 there was a second babyboom which lasted some 12-13 years.
\qpar

Before we leave this topic a last remark is in order.
There has been an ongoing debate about the amount of excess deaths 
during the years 1958-1961. Unfortunately, 
at this point we have no way to make reliable estimates based
on population pyramids%
\qfoot{In the next section we explain why this is a more
difficult problem.}%
. 
This is an ongoing work however and
we hope that in the future it will be possible to solve this question.

\qA{Another cause of birth reduction}

Just to show that the factors mentioned
previously are not the only causes
of sudden reductions in birth numbers we describe here
a very different cause.
In Fig. 4 there are two events labelled ``Fire Horse'' which
are marked by sudden, short-lived birth reductions. These events
occurred in 1906 and 1966 respectively. The expression ``Fire Horse'' 
refers to the Chinese calendar. This calendar comprises
12 different
animals such as ``Horse'' and 5 different symbols such a ``Fire''.
This means that a specific animal together with a specific symbol will
occur every $ 12\times 5=60 $ years. In Japan there
is a strong belief%
\qfoot{Although the same calendar is used there is
no similar belief in China.}
that a daughter born in a Fire Horse year will bring
ill-being and suffering to her family. As a result some 10\%
of the couples preferred to postpone a possible birth to a 
later year.  The fact that this reduction occurred twice and exactly 
in the expected years suggests that this explanation
is indeed the right one.
\qpar

Previously we mentioned factors which affect people in their
material living conditions. The Fire Horse events show that
social beliefs can have an equally strong impact on births.

\qI{Simple versus difficult cases}

In the previous section we have seen that abrupt changes in birth
numbers sharply affect the shape of population pyramids.
Why is this so? The reason is very simple.
For any age-group the birth year represents a single year.
So any short-lived event which affects births will leave its
mark on only one or two steps of a pyramid. On the contrary,
an event such as an earthquake or a disease will 
affect all ages and reduce many
steps of a pyramid. As a result, it will be difficult to
distinguish this effect from a multi-year trend in birth rate.
\qpar

In addition the effect of special events on birth rates
can lead to reductions as large as 50\%. 
On the contrary, 
even major wars hardly ever kill more than 10\% 
of an age group.
For instance, over one million French soldiers were killed
on the battle fields of World War I but these deaths were distributed
over several age groups with the result that in each age-group
the reduction did not exceed 15\%.
\qpar

In short, any event that (i) is concentrated in time 
(ii) affects only one or two age-groups 
and (iii) affects them strongly (effect larger than 10\%)
will be easy to analyze through population pyramids.
On the contrary, episodes which last several years and
affect a whole 
population from young to elderly will have only a ``diluted'' 
impact on age-groups and therefore will be hard to detect and
even more difficult to analyze.

\qI{Population pyramids of North and South Korea}

In the previous sections we have shown two things:
\qdec{(i) That population pyramids can be used to estimate birth rates.\qL
(ii) That a dip in birth rate provides a signature of 
hardship whether due to war or to other factors.}
\qpar

In the present section we wish to apply these clues to a specific 
case, namely a comparison between the population pyramids of North
and South Korea (Fig. 7). 

\begin{figure}[tb]
    \centerline{\psfig{width=16cm,figure=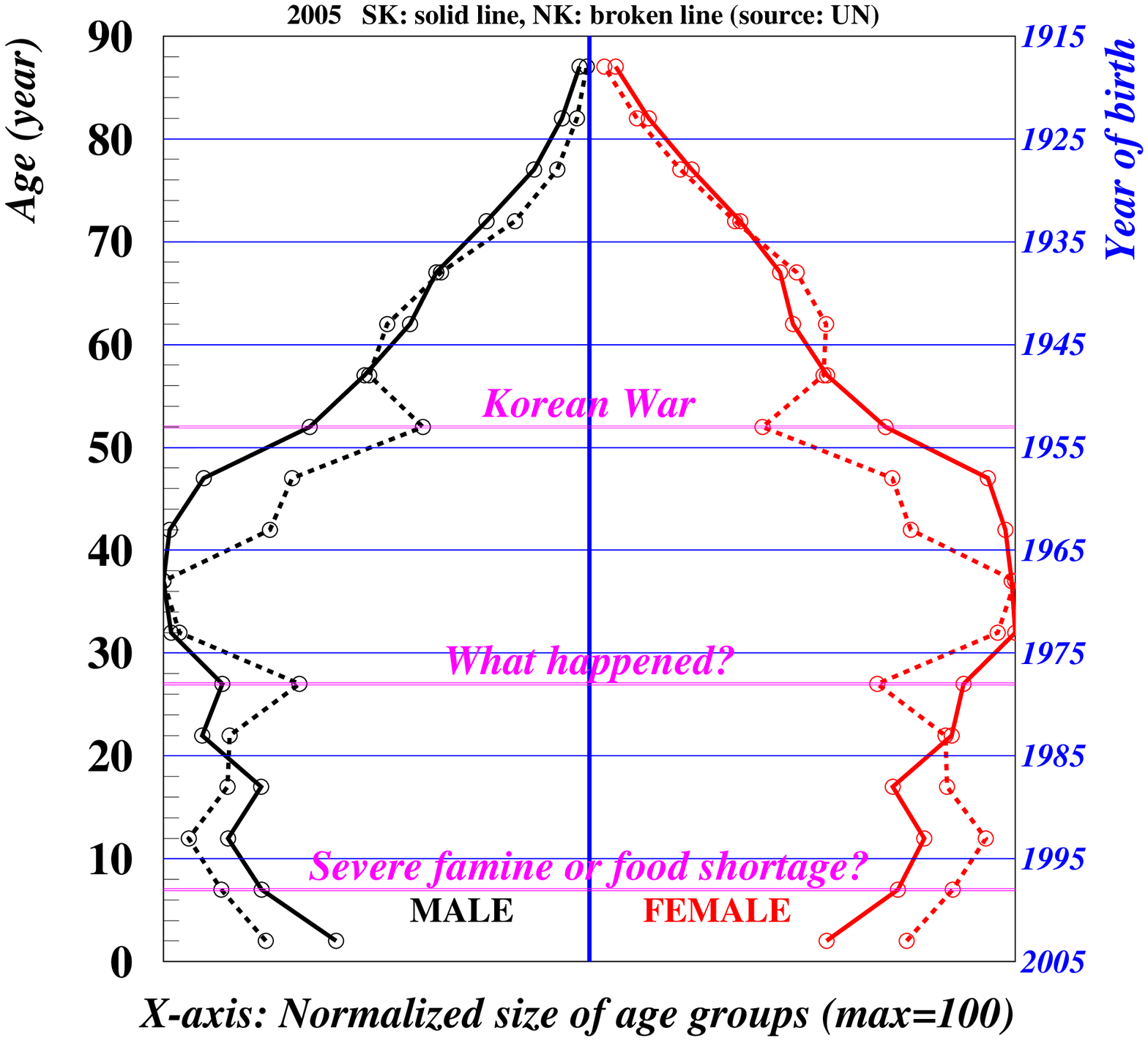}}
\qleg{Fig.\qhu 7:\qhv Comparison of the population
pyramids of South and North Korea.}
{The pyramid represents the age groups in 2005.
The solid line is for South Korea, the broken line for
North Korea.
The
comparison raises the question of what caused the dip of
1978. Two factors seem to have played a role:
(i) The fact that
the reduced age groups of the Korean War reached marriage age 
(ii) The birth-control program launched by the North Korean government
possibly following the example of the one-child policy which was introduced
in China around this year.}
{Source: United Nations: World Population Prospects, 2008 revision.}
\end{figure}

 \qA{Similarities}
First of all, by comparing the broad shapes of the two
pyramids we see that they look fairly similar. From top to
bottom (that is to say from 1915 to 2005) one can mention 
the following features.
\qbu For women of
old age (over 65) the slopes are almost the same which points to
similar conditions in caring for the elderly.  The case of men
seems different but this comes from a depletion of these age groups
before World War II (as will be seen below).
\qbu In contrast to the case of European countries or Japan 
there was no babyboom in North and South
Korea after the end of World War II%
\qfoot{A babyboom can be defined by the fact that 
in the years after the war the birth rate is higher than it had been
before the war. France provides a spectacular illustration.
Between 1946 and 1969 the birth rate was on average 50\% higher
than between 1930-1939. In Japan birth rates in 1947-1950 were
25\% higher than before the war but after 1950 they quickly came
back to their pre-war level. Fig. 7 suggests that there was
a small increase in births in South Korea between 1945 and 1950
but the real babyboom came after 1953.}%
.
\qbu In both countries a babyboom started at
the end of the Korean War and lasted until 1970.
\qbu Both pyramids show a narrow base
which indicates a reduction in birth rates after 1975. 

\qA{Differences}
One can note the following differences.
\qbu From the 1965 pyramid it can be seen that
for North Korean people born between 1900 and 1935 the ratio 
male/female is much lower than for South Korean people. It becomes
as low as $ 0.6 $  whereas in South Korea it is almost $ 1 $. 
At this point we have no explanation.
\qbu Whereas the Korean War did not much affect the birth rate
in South Korea it strongly affected the birth rate in North Korea.
\qbu The North Korean pyramid shows a substantial indentation
around 1978. For South Korea there is also a reduction but it is
more a trend than a sudden event. This indentation can be
explained by two factors. (i) It occurs some 25 years after the 
Korean War which means that the reduced age groups due to 
the war will have fewer children than the normal corresponding
age groups in South Korea. However, because of the dispersion 
in the age at marriage this effect should be less abrupt than in
1953 which means that there was probably a second factor. Indeed,
in these years (and probably following the example of China)
the North Korean government started a policy of birth control.
More details can be found in the article ``Changes in population
of North Korea''.

\qA{Was there a severe famine in North Korea?}

Although this point does not originate from a comparison of the two
pyramids one can hardly avoid this question because it has
generated a vivid debate in recent years and even up to now.
On the one hand, there are some Western experts who claim 
that around 1998
there was a severe famine in North Korea which caused between 600,000
and 4 millions deaths. On the other hand one has the
thesis that there was indeed
a food shortage but no real famine with many deaths.
What is the truth?
\qpar

We said that it is not the examination of the pyramid which raises
this question. Indeed,
the pyramid show nothing special in 1998.
According to our previous hardship criterion a famine should have produced
a marked indentation, at least of same magnitude as the one in 1978. 
But nothing of the sort can be observed.
\qpar

As a matter of fact, if one believes that the population data
transmitted to the United Nations by the North Korean 
government are basically correct, the famine thesis does not hold.
This follows not only from the population pyramid but also, more
simply, from the
examination of the total population. During the 1990s and 2000s
in normal years the natural population increase (due to the 
difference between birth and death 
rates) was about $ 15-7=8 $ per thousand. With a population of
about 20 millions one gets an annual surplus of 160,000. 
Now, even if one takes 
the lowest famine death toll estimate, 
namely 600,000, and if one assumes that this toll
was spread over 3 years, one gets an annual population change of
$ 160,000-600,000/3 = -40,000 $ which means that the population
should have been decreasing over 3 years.
This is in contradiction
with the data about the North Korean population%
\qfoot{See for instance on Wikipedia the file
Korea-North-demography.png}%
.
The curve which shows the evolution of the population
between 1961 and 2003 does not reveal any annual fall.
\qpar

In short, the two claims are incompatible.
One of them must be wrong. 
Because it would lead us too far away from the main
topic of this article we will not here try to discuss the validity
of the two theses. More details can be found in the longer version
of the paper which is available on the author's homepage. 
In particular, it will be seen that the internal documents of 
two organizations which should be well informed on this matter, 
namely the
US State Department and the United Nations Food and Agriculture
Organization, hardly ever mention a famine in North Korea.
They mention a chronic food shortage but they do {\it not} mention
people starving to death. Beyond this debate, what is perhaps the
most important point
from a scientific perspective is to realize that 
the thesis of a famine which caused 2 million deaths is not
as well established as the broad coverage it has received in
western media would lead us to think%
\qfoot{An example chosen at random among many similar sentences
reads as follows. ``The exodus of North Koreans to Jilin and Liaoning Provinces [North of China]
began in earnest in the waves of famine that struck North Korea in the mid-1990's, killing as many as two million people'' (New
York Times 24 March 2005).
It can be noted that the 2 million death toll is presented here as
a {\it fact} not an estimate or a conjecture. 
Incidentally, there was also an
exodus of Polish people to the UK, Ireland and Iceland 
in the 2000s without any famine in Poland; they were just 
seeking better wages.}%
.

\qI{Conclusion and perspectives}

In this article we tried to convince the reader that population
pyramids are an effective tool for exploring
the demographic facet of social phenomena. 
As the purpose of the paper
is to provide an overall introduction it focused on 
basic principles and some illustrations. 
Many questions still need to be investigated more closely.
For instance in the case of a war what is the key-factor?
Is it the death of soldiers, the occupation of parts of
the country by foreign troops, the destruction of 
cities by air raids or some other factor.

\qA{As in physics one can study each factor separately}

In contrast to many other questions in the social sciences
for which we cannot set up many experiments, here we can.
We are in the same position as the astrophysicists 
mentioned at the beginning of the paper.
We have enough cases at our disposal 
to isolate one factor after another. For instance, 
if we wish to study the effect of air raids, we
can investigate the cases of Germany or Japan in 1945, of
North Korea in the Korean War, of North Vietnam in the Vietnam War,
of Iraq in the First and Second Gulf Wars. 
Similarly, for occupation cases there are many specific episodes.
\qpar

In fact, we are even in a better position than astrophysicists
because we can first test our methodologies on {\it known}
cases. An example will explain what we mean.
By using the population
pyramids of Japan for the years 1898, 1903 and 1908 it is 
possible to estimate the number of excess deaths due to the
Russo-Japanese War of 1905. This provides a useful test of the
methodology because it is possible to compare such excess-deaths
to what historians tell us about the toll of the war.
\qpar

\qA{Research agenda}
This example suggests  a 4-step research agenda.\qL
(i) First one develops a new measurement method (ii) Then one tests
it on several ``known'' cases.
(iii) If the death estimates provided by the method agree with those
given by some reliable sources, the method will be validated.
(iv) Once validated, it
can be used to explore cases for which there
are no reliable data or for which there are conflicting data. 

\qA{How to find new measurement methods?}

For discovering new measurement methods (the first step in our
previous agenda) the strategy that we suggest
is inspired from what physicists do. To make this point clearer
it may be useful to recall a well-known example.
\qbu Light rays usually move along straight lines. This
is what can be called the basic rule.
\qbu However, in some cases they do {\it not} move along
straight lines. This signals that some ``special event''
is taking place. For instance, the basic rule will not hold
in a substance whose
index of refraction is not uniform or
in the strong gravitational field that exists in the
vicinity of the Sun. In such cases the trajectory of the light-ray
is bent.
\qbu By measuring the angle of deviation of the light-rays, it is
possible to get information about the ``special events''. 
Thus, in the two previous cases
one can determine the gradient of the refraction index or
the strength of the gravitational field.
\qpar

Similarly, our strategy will comprise the following steps.
{\color{blue} \qbu} First, one must analyze the process 
of aging and discover some ``basic rule''. 
As an example of such a rule, one can mention
Gompertz' law which says that after
the age of 40 the probability of death doubles every 8 years.
{\color{blue} \qbu} When population pyramids show 
a deviation away from the basic
rule this tells us that a special event has occurred. 
{\color{blue} \qbu} By comparing actual data to what the
rule would led us to expect, one can determine the characteristics
of the ``special event''.

\appendix

\qI{Appendix A. Sources for census data}

In order to build population pyramids by single year of age 
for the different provinces of a country,
one needs census data about age and place of residence.
Moreover, one would like to find such statistics for as 
many countries and as many dates as possible. 
Where can one find such information? There are several
possible sources. Thanks to the Internet all of them
are fairly easily available.

\qA{National statistical organizations}

For census data the primary sources are the statistical agencies
of each country. Although more and more countries publish 
their statistical yearbook in bilingual form (national language
+ English) the more detailed statistical data that we need
are usually not published in bilingual form. In other
words to get access to such Japanese data (for instance) one needs 
some knowledge of the Japanese language.
\qpar

Fortunately, there are some websites which provide the information
that we need for {\it many} countries. 
In what follows we restrict ourselves to such sources which are
freely available on the Internet%
\qfoot{We do not give the addresses of the websites because
usually Internet addresses have only a short life-time. However, all
these websites can be easily located through their titles with
the help of a search engine.}%
.

\qA{The IPUMS websites}

IPUMS is an acronym which means ``Integrated Public Use Microdata Series''.
The word ``microdata'' 
means that this website provides {\it individual} census data 
for representative samples (usually 1\% samples)
of the whole population of a country. There are two IPUMS
websites: ``IPUMS USA'' 
gives data for American censuses from 1850 to 2000. ``IPUMS International''
covers several countries%
\qfoot{E.g. Argentina, Brazil, Chile, China, France, Egypt, India,
Iraq, Kenya, Mexico, Pakistan, Palestine, Philippines, Puerto Rico,
South Africa, Spain, Switzerland, Thailand, United Kingdom,
Venezuela, Vietnam. It can be noted that some important countries
are missing, e.g. Germany, Indonesia, Japan, Korea, Russia.
Except for a few cases the dates are posterior to 1970.}%
.
Through these data bases one can build population pyramids 
at regional level (provinces for China, canton for Switzerland, and
so on).

\qA{UN database of populations by single years of age.}

The United Nation provides population data by sex and single year
of age for all almost all countries and at various dates.
These data can be used very easily to build population
pyramids at country level.

\qA{UN database of populations by 5-year age groups (1950-2010)}

The Population Division of the United Nations 
periodically publishes
a dataset entitled: ``World Population Prospects''. 
The 2008 Revision was released on 1 September 2010. Among other
things, this database gives the population of all countries by sex and
5-year age groups. 
\qpar

It should be noted that,
strictly speaking these data are {\it not}
census data. This is clear from the fact that the tables are given
for the same years (1950, 1955, 1960, 1965, $ \ldots $) in each country
whereas actual censuses take place in different 
years depending on the country. In other words, the data given
in this database are estimates computed by the Population Division.
\qpar
There are two sections about ``Sources'' and ``Assumptions''
in which  official sources are listed.
However, this list mentions only the most recent censuses.
For instance in the case of Korea (North and South) no indication
is given about the sources which were used for the estimates
of 1950 and 1955.
Moreover, the ``Assumptions'' section
does not say precisely 
how the interpolations (or extrapolations)
have been performed. The assertion which is made that
``population data from all sources were evaluated 
for completeness, accuracy and consistency, and 
adjusted as necessary'' actually means 
that one must trust the experts who produced the
estimates. We do not intend to say that the estimates
are not good but rather that one cannot know how good they are.
\qpar

In addition to the previous multi-national websites
one can also mention a very convenient website
which gives historical data for Japan.

\qA{Historical Statistics of Japan}

The statistical series on the
website ``Historical Statistics of Japan'' are
published by the Japanese Ministry of Internal Affairs and
Communication. The population data 
give the distribution by single years of age
for all censuses held in Japan between 1884 and 2000.
\vskip 4mm

{\bf Acknowledgements} \quad The author would like to thank
the organizers of the ``4th Chinese-Europe Summer School on
Complexity Sciences'' held in Shanghai in August 2010. By
providing a large and interested audience for the present topic,
this meeting boosted the analysis of population pyramids.
Many thanks also to Professors Hawoong Jeong and Beom Jun Kim 
for their interest and useful discussions especially about
the demographic history of Korea.
\qpar

The present paper was published in the APCTP Bulletin of 2010
(Roehner 2010, p. 13-25).

\vskip 8mm

{\bf \large References}
\vskip 4mm

\qparr
Barab\'asi (A.-L.) 2010: Bursts. The hidden pattern behind 
everything we do. Dutton Adult.

\qparr
Brownstein (C.A.), Brownstein (J.S.) 2008: Estimating excess mortality
in post-invasion Iraq. 
New England Journal of Medicine 358,5,445-447.

\qparr
Change in population of North Korea and policy implications for
health and welfare. [in Korean]\qL
See in particular tables 2 and 3 (p. 38-40) 

\qparr
Dumont (R.) 1964: Les communes populaires rurales chinoises
[The people's communes in rural China].
Politique \'Etrang\`ere 29,4,380-397.\qL
[available online]

\qparr
Liesner (T.) 1989: One hundred years of economic statistics. Facts
on File, New York.

\qparr
Roehner (B. M.) 1990: Corr\'elations entre fluctuations des prix
et fluctuations d\'emographiques. France XIXe si\`ecle. 
[Correlations between the fluctuations of wheat
prices and of vital population numbers in
nineteenth century France.]\qL 
Population 45,2,299-326.

\qparr
Roehner (B.M.) 2010: How can population pyramids be used to
explore the past?
Asia Pacific Center for Theoretical Physics (APCTP) Bulletin,
No 25-26,13-25, January-December 2010.

\vskip 5mm

Bertrand M. Roehner
\qbu Permanent address: Institute for Theoretical and High Energy Physics,
LPTHE, 4 place Jussieu, F-75005 Paris, France. \qL
E-mail: roehner@lpthe.jussieu.fr, Phone: 33 1 44 27 39 16\qL
Homepage: http://www.lpthe.jussieu.fr/$ \sim $roehner
\qbu July-December 2010:
Department of Systems Science, Beijing Normal University\qL
19 Xiejiekouwai Street, Beijing 100875, China. 

\ifnum\count101=1
\end{document}
\fi